\documentclass[a4paper,11pt]{article}
\usepackage[a4paper]{geometry}
\usepackage[T1]{fontenc}
\usepackage{lmodern}
\usepackage{graphicx}
\usepackage{color}
\usepackage{amsmath}
\usepackage{amsfonts}
\usepackage{bbm}
\usepackage{array}

\def\diffd{\mathrm d}
\def\Z{\mathbb Z}

\def\one{\mathbbm{1}}

\def\cst{\hbox{Cst}}
\def\one{\mathbbm 1}

\author{\'Eric Brunet \& Bernard Derrida\footnote{\texttt{Eric.Brunet@lps.ens.fr}, \texttt{Bernard.Derrida@lps.ens.fr}.
\'E.B.\@ and B.D.: Sorbonne Universit\'es, UPMC Univ Paris 06, CNRS, UMR
8550, LPS-ENS, F-75005, Paris France.
B.D.: Coll\`ege de France, F-75005, Paris France.
}}

\title{An exactly solvable travelling wave equation in the Fisher-KPP class}
\begin{document}
\maketitle

\begin{abstract}
For a simple one dimensional lattice version of a travelling wave
equation, we obtain an exact relation between the initial condition and the
position of the front at any later time. This exact relation
takes the form of an inverse problem: given the times $t_n$ at which the
travelling wave reaches the positions $n$, one can deduce the initial
profile. We show, by means of complex analysis, that a number of known
properties of travelling wave equations in the Fisher-KPP class can be
recovered, in particular Bramson's shifts of the positions. We also recover
and generalize Ebert-van Saarloos' corrections depending on the initial
condition.
\end{abstract}

\section{Introduction}
%=====================

The study of the solutions of partial differential
equations describing a moving interface from a stable to an
unstable medium is a classical subject\cite{AronsonWeinberger.75,
McKean.75, Kametaka.1976, Bramson.78, Bramson.83}  in mathematics,
theoretical physics and biology \cite{DerridaSpohn.88,Murray.2002,
MeersonVilenkinSasorov.2013,Munier.14}. The prototype of such
equations is the Fisher-KPP equation (after Fisher \cite{Fisher.37} and
Kolmogorov-Petrovskii-Piskunov \cite{KPP.37})
\begin{equation}
\frac{\partial u}{\partial t} = \frac{\partial^2 u}{\partial x^2} + f(u),
\label{FKPP}
\end{equation}
where the field $u$ satisfies $0 \le u(x,t) \le 1$ and where $f(u) \ge
0$. The unstable medium corresponds to $u=0$ (i.e.\@ $f(0)=0$ and $f'(0) >0
$) and the stable one to $u=1$ (i.e.\@ $f(1)=0$ and $f'(1)<0$). 

One can show that equations of type \eqref{FKPP} exhibit a continuous
family $W_v$ of travelling wave solutions
\begin{equation}
u(x,t) = W_v(x-v t) \label{TW}
\end{equation}
indexed by their velocities $v$. Explicit expressions of the travelling
waves are in general not known except for particular velocities
\cite{MaFuchssteiner.1996}. The best known example, due to Ablowitz
and Zeppetella \cite{AblowitzZeppetella.1979}, is $u= \big[1+ c \exp[(x-v t)
/\sqrt{6} ]\big]^{-2}$ for the Fisher-KPP equation~\eqref{FKPP} with
$f(u)=u-u^2$ and $v=5/\sqrt6$.

Apart from describing the shapes of these travelling wave solutions
\eqref{TW}, a central question is to understand how the long time behavior
of the solutions of \eqref{FKPP} depends on the initial condition $u(x,0)$.
In general this asymptotic regime is controlled by the rate of the
exponential decay of this initial condition. A brief review of the
properties of the travelling wave solutions of \eqref{FKPP} and on the way
the position and the asymptotic velocity of the solution depend on the
initial condition is given in Section~\ref{S:recall}.

In the present paper we study a simple one dimensional lattice version of
a travelling wave equation. In this lattice version we associate to each
lattice site $n\in{\mathbb Z}$ a positive number $h_n(t)$ which plays the role of the
field $u(x,t)$ and these $h_n(t)$ evolve according to
\begin{equation}
\frac{\diffd h_n(t)}{\diffd t} =
\begin{cases}
 a h_{n-1}(t) + h_n(t) & \text{if $0 \le h_n(t) <1$},\\
 0 & \text{if $h_n(t)\ge 1$}.
\end{cases}
\label{evolution}
\end{equation}
We see that the evolution of $h_n(t)$ is linear except for the
saturation at $h_n(t)=1$ which is the only non-linearity in the problem.
This saturation simply means that whenever $h_n(t)$ reaches the value $1$,
it keeps this value forever. The evolution~\eqref{evolution} therefore
combines linear growth, spreading (or diffusion) because of the coupling
between neighboring sites, and saturation, very much like in Fisher-KPP
equation~\eqref{FKPP}. 

The aim of this paper is to show that the evolution~\eqref{evolution} leads
to behaviors very similar to those expected for the usual Fisher-KPP
equation~\eqref{FKPP}. Moreover a number of properties of the solutions
of~\eqref{evolution} are easier to determine than for the original Fisher-KPP
equation~\eqref{FKPP}. Our approach is essentially based on the exact
relation~\eqref{key-formula} derived in Section~\ref{S:key} which relates
the times $t_n$ at which $h_n(t)$ reaches~1 for the first time to
the initial condition $h_n(0)$. We show in Section~\ref{S:TW} that from
\eqref{key-formula} one can obtain a precise description of the shape
of the travelling wave solutions, in particular explicit formulas for their
asymptotic decay. We also show in Section~\ref{S:position}
that~\eqref{evolution} shares with the Fisher-KPP equation most of the
properties expected for the dependence of the position of the front on the
initial condition. Our results  are summarized in
Section~\ref{sec:summary}.

\section{Some known properties of the Fisher-KPP class}
%======================================================
\label{S:recall}

In this section we briefly recall some properties of the Fisher-KPP equation.

\subsection{The travelling waves }
%---------------------------------

For the Fisher-KPP equation \eqref{FKPP} the shape $W_v(x)$ of the travelling
wave \eqref{TW} satisfies an ordinary differential equation
\begin{equation}
\label{ode}
W_v'' + v W_v' + f(W_v)=0
\end{equation}
with the boundary conditions $W_v(-\infty)=1$ and $W_v(+\infty)=0$. By
linearizing \eqref{ode} for small $W_v$ (when $x$ is large),
\begin{equation}
\label{lode}
W_v'' + v W_v' + f'(0) W_v=0,
\end{equation}
one can see that, generically, $W_v(x)$ vanishes exponentially as $x \to
\infty$
\begin{equation}
W_v(x) \sim e^{-\gamma x},
\label{expo-decay}
\end{equation}
with $\gamma$ related to the speed $v$ of the travelling wave by
\begin{equation}
v(\gamma) = \gamma + \frac{f'(0)}{ \gamma}.
\label{v-KPP}
\end{equation}

This relation shows that depending on $v$, the rate $\gamma$ of the
exponential decay is either real or complex, and these two regimes are
separated by a critical velocity $v_c$ where $v(\gamma)$ is minimum
\begin{equation}
\label{vc}
 v_c= v(\gamma_c) \qquad\text{where}\qquad v'(\gamma_c) =0.
\end{equation}
With $v(\gamma)$ given by \eqref{v-KPP}, one gets $v_c=2
\gamma_c$ and $\gamma_c=\sqrt{f'(0)}$. Under certain conditions on the
function $f(u)$ (such as $0\le f(u)\le u f'(0)$ for all $u$  see
\cite{Bramson.78,Bramson.83,vanSaarloos.03} and references therein), it is
known that:
\begin{itemize}
\item For $0<v<v_c$, the solutions $\gamma$ of the equation $v(\gamma)=v$ are
complex. The corresponding travelling waves solutions of~\eqref{ode}
oscillate around $0$ while decaying as $x \to \infty$. 
\item For $v>v_c$, the travelling wave is monotonically decreasing and
decays for large $x$ as 
\begin{equation}
 W_v(x) \simeq A\,e^{-\gamma_1 x}\qquad\text{with $A>0$},
\label{decay1}
\end{equation}
where $\gamma_1$ is the smallest solution of $v(\gamma)=v$.
\item For $v=v_c$, the equation $v(\gamma)=v_c$ has a double root
$\gamma_c$ and the travelling wave is monotonically decreasing and decays
for large~$x$ as 
\begin{equation}
W_{v_c}(x) \simeq A \, x\, e^{-\gamma_c x}
 \qquad\text{with $A>0$}.
\label{decay2}
\end{equation}
\end{itemize}
\textit{Remark:} The facts~\eqref{decay1} and~\eqref{decay2} for $v \ge
v_c$ are not obvious and cannot be understood from the linearized
equation~\eqref{lode} only. These are properties of the full non-linear
equation~\eqref{ode}, which can be proved under known conditions on the
non-linearity $f(u)$ (such as $0\le f(u) \le u f'(0)$). Fronts which
satisfy these properties are called
\emph{pulled fronts}.

For well tuned non-linearities (which fail to satisfy
these conditions), travelling waves for $v\ge v_c$ might not be monotone
and the asymptotics \eqref{decay1} and \eqref{decay2} might be modified;
for instance in~\eqref{decay1}, depending on the value of $v$, one could
have $A<0$ or a decay in $\exp(-\gamma_2 x)$ where $\gamma_2$ is the
largest solution of $v(\gamma)=v$. Rather than~\eqref{decay2}, one could have
$A\exp(-\gamma_c x)$ without the $x$ prefactor. In all these cases,
the front equation is then
said to be \textit{pushed} \cite{BenguriaDepassier.1996,vanSaarloos.03}.

\subsection{The selection of the velocity}
%-----------------------------------------

The travelling waves $W_v$ solutions of~\eqref{ode} move at a constant
speed with a time independent shape. For general initial conditions
$u(x,0)$, the shape of the solution is time-dependent and the question
of the selection of the speed is to predict the asymptotic shape and
velocity of the solution $u(x,t)$ in the long time limit. For
initial profiles decreasing from  $u(-\infty,0)=1$ to $u(+\infty,0)=0$ it
is known since the works of Bramson
\cite{Bramson.78, Bramson.83, vanSaarloos.03,
HamelNolenRoquejoffreRyzhik.2013, Munier.14} 
under which conditions the shape of the solution
$u(x,t)$ converges to a travelling wave $W_v$ solution of \eqref{ode} in
the sense that one can find a displacement $X_t $ such that
\begin{equation}
u(X_t+x,t) \to W_v(x),\quad\text{with }\frac{X_t}t\to v.
\label{cvg}
\end{equation}

In particular it is known that 
if the initial condition $u(x,0)$ satisfies for large~$x$: 
\begin{itemize}
\item $u(x,0)\sim e^{-\gamma x}$ with $0<\gamma<\gamma_c$,\\
then the asymptotic velocity is $v(\gamma)$, the asymptotic shape is
$W_{v(\gamma)}$ and 
\begin{equation}
X_t=v(\gamma) t + \cst. 
\label{Xt1}
\end{equation}

\item $u(x,0)\ll  x^\alpha e^{-\gamma_c x}$ for some
$\alpha < -2$ (in particular for step initial conditions),
\\the asymptotic velocity is
$v_c=v(\gamma_c)$, the asymptotic shape is $ W_{v_c}$ and 
\begin{equation}
X_t=v_c t -\frac3{2\gamma_c}\ln t+ \cst .
\label{Xt2}
\end{equation}

\item $u(x,0)\sim x^\alpha e^{-\gamma_c x}$ with $\alpha \ge -2$,\\
the asymptotic velocity $v_c$ and shape $W_{v_c}$ are the same as in the
previous case but the logarithmic correction to the position $X_t$ is
modified:
\begin{align}
X_t &=v_c t -\frac{1-\alpha}{2\gamma_c}\ln t+ \cst
&&\text{for $\alpha>-2$,} \label{Xt3}\\
X_t &=v_c t -\frac{3}{2\gamma_c}\ln t+
\frac1{\gamma_c}\ln \ln t+ 
 \cst &&\text{for $\alpha=-2$.}
\label{Xt4}
\end{align}
\end{itemize}
(Initial conditions decaying too slowly would not lead to a travelling
wave.)

Notice that the solutions $W_v$ of \eqref{ode} can always be translated
along the $x$ axis, so the ``$\cst$'' in (\ref{Xt1}-\ref{Xt4})
depends on the particular solution of~\eqref{ode} that was chosen. It is
often convenient to single out one particular solution $W_v$ of \eqref{ode}:
for example one may select the
solution such that $W_v(0)=1/2$ or such that $\int x W_v'(x)
\,\diffd x =0$. Once a particular prescription for 
$W_v$ is chosen, the ``$\cst$'' in the equations above is well defined. It
can be computed in some cases such as~\eqref{Xt1}, but its analytic
expression is not
known in some other cases such as~\eqref{Xt2}.

\subsection{Vanishing corrections}

The convergence property~\eqref{cvg} does not allow to define the
displacement~$X_t$ to better than a constant: if $X_t$ 
satisfies~\eqref{cvg}, then $X_t+o(1)$ also satisfies~\eqref{cvg}.
It is however quite natural to choose a particular~$X_t$, which one might
call the position of the front. A possible choice could be
\begin{equation}
u(X_t,t) = c,
\label{Xtc}
\end{equation}
where $c\in(0,1)$ is a fixed given number. Another possible choice would be
to interpret $-\partial u/\partial x$ as a probability density and pick
$X_t$ as its expectation:
\begin{equation}
X_t=-\int\diffd x\, x \frac{\partial u}{\partial x}.
\label{Xte}
\end{equation}
Either definition~\eqref{Xtc} or~\eqref{Xte} gives a position $X_t$ which
satisfies~\eqref{cvg}. With such a precise definition
for~$X_t$ as~\eqref{Xtc} or~\eqref{Xte}, it makes sense to try to
improve on (\ref{Xt1}-\ref{Xt4}) and determine higher order corrections.
Ebert and van Saarloos \cite{EbertvanSaarloos.00,
MuellerMunier.2014} have claimed that for steep
enough initial conditions, the first correction to~\eqref{Xt2}
is of order $t^{-1/2}$ and is universal: it depends neither on the initial
condition, nor on the choice of~\eqref{Xtc} or~\eqref{Xte}, nor on the value~$c$
in~\eqref{Xtc}, nor on the non-linearities. They found that
\begin{equation}
 X_t = v_c t - \frac3{2 \gamma_c } \ln t + \cst -  3\sqrt{\frac{2 \pi}
{ \gamma_c^{5} v''(\gamma_c) }}\, t^{-1/2} + \cdots. 
\label{EbSa}
\end{equation}

\subsection{The Fisher-KPP class}
%--------------------------------

The main ingredients of the Fisher-KPP equation~\eqref{FKPP} which lead to
travelling waves and fronts converging to those travelling waves
are a diffusive term, a growth term and a saturation
term. There exist many equations with the same ingredients which share
the above properties (\ref{vc}-\ref{EbSa}) of the Fisher-KPP
equation: the equation satisfied by the travelling waves~\eqref{ode}, the
dispersion relation \eqref{v-KPP} and the values of $\gamma_c$ and $v_c$
are modified, but everything else remains the same.

To give an example which appears in the problem of directed
polymers on a tree~\cite{DerridaSpohn.88}, let us consider an evolution
equation of the type
\begin{equation}
G(x,t+1)=\int G(x+\epsilon)^B  \rho(\epsilon) \, \diffd \epsilon.
\label{tree}
\end{equation}
(In the directed polymers context, $B$ is the branching ratio on the tree
and $\rho(\epsilon)$ is the
distribution of the random energies associated to edges of the tree).
Then $u(x,t)=1-G(x,t)$ satisfies a discrete time evolution equation with an
unstable uniform solution $u=0$ and a stable one $u=1$ as in \eqref{FKPP}.
Even though~\eqref{FKPP} is continuous in time while \eqref{tree} is
discrete, they have similar properties: travelling waves
for~\eqref{tree} are solutions of 
\begin{equation}
W_v(x-v) = \int W_v(x+\epsilon)^B  \rho(\epsilon) \, \diffd \epsilon
\end{equation}
instead of \eqref{ode}. By linearizing the evolution of $G(x,t)$ around
the unstable uniform solution $G=1$ and by looking for travelling wave
solutions  of this linearized equation of the form $1-G(x,t) \sim \exp[
-\gamma(x - v(\gamma) t )]$, one gets a new dispersion equation which
replaces \eqref{v-KPP}:
\begin{equation}
v(\gamma) = \frac{1}{\gamma} \ln \left[ B \int e^{\gamma \epsilon}
\rho(\epsilon) \,\diffd \epsilon \right],
\label{new-v}
\end{equation}
but all the above behaviors (\ref{vc}-\ref{EbSa}) remain
valid with $v_c$ and $\gamma_c$ computed from \eqref{new-v} and~\eqref{vc}.
For example, for $B=2$ and
a uniform $\rho(\epsilon)$ on the unit interval (i.e.\@ $\rho(\epsilon)=1$
for $0< \epsilon <1$ and $\rho(\epsilon)=0$ elsewhere) one gets,
$v(\gamma)=\frac1\gamma\ln\big[\frac2\gamma(e^\gamma-1)\big]$ which leads
to $v_c\simeq0.815172$ and $\gamma_c\simeq5.26208$.

Example \eqref{tree} is a front equation where time is
discrete. One could also consider travelling wave equations where space is discrete, say
$x\in \Z$. For instance, one could discretize the Laplacian in~\eqref{FKPP}
or take \eqref{tree} with a distribution $\rho(\epsilon)$
concentrated on integer values of $\epsilon$. When space is discrete,
special care should be taken: it is clear from~\eqref{TW} that while the
front $u(x,t)$ lives on the lattice, the travelling wave $W_v(x)$ is
defined for all real values~$x$, and even when~\eqref{TW} holds,  the shape
of the front $W_v(x-v t)$ measured on the lattice evolves periodically
in time with a period $1/v$. Furthermore, the convergence~\eqref{cvg} no
longer makes any sense. One can still try to define a specific position of
$X_t$ by something like the following generalization of~\eqref{Xte}:
\begin{equation}
X_t=\sum_{x\in\Z} x\big[u(x,t)-u(x+1,t)\big],
\label{Xte2}
\end{equation}
but with such a definition, even if the front is given by the travelling
wave $W_v(x-vt)$, the difference $X_t-vt$ is no longer constant but becomes
a periodic function in time because the shape of the front on the lattice
evolves also periodically. Similarly, in the discrete space case, the $\cst$
term in all the asymptotics~(\ref{Xt1}-\ref{Xt4}) is in general replaced by
a periodic function of time.

An alternative way to locate the front when time is continuous and space is
discrete is to invert the roles of $x$ and $t$: instead of defining $X_t$
by $u(X_t,t)=c$
as in~\eqref{Xtc}, one can define $t_x$ as the first time when the front at
a given position $x$ reaches a certain level~$c$:
\begin{equation}
u(x,t_x)=c.
\label{txc}
\end{equation}
Note that when time and space are continuous, the functions $X_t$ and $t_x$ are
reciprocal and one can write (\ref{Xt1}-\ref{Xt4}) as
\begin{align}
\label{tx1}
t_x &=\frac{x}{v(\gamma)}+\cst',&&\text{for $u(x,0)\sim e^{-\gamma x}$ with
$0<\gamma<\gamma_c$,}\\[1ex]
t_x &=\frac{x}{v_c}+\frac3{2\gamma_c v_c}\ln x+\cst',&&\text{for
$u(x,0)\ll x^\alpha e^{-\gamma_c x}$ for some $\alpha<-2$,}\\[1ex]
t_x &=\frac{x}{v_c}+\frac{1-\alpha}{2\gamma_c
v_c}\ln x+\cst',&&\text{for $u(x,0)\sim x^\alpha e^{-\gamma_c x}$ for 
$\alpha>-2$,}\label{tx3}\\[1ex]
t_x &=\frac{x}{v_c}+\frac{3}{2\gamma_c
v_c}\ln x-\frac1{\gamma_cv_c}\ln\ln x+\cst',&&\text{for $u(x,0)\sim x^{-2} e^{-\gamma_c x}$,}
\label{tx4}
\end{align}
and, for steep enough initial conditions, one can write \eqref{EbSa} as
\begin{equation}
t_x=\frac x{v_c}
+\frac
1{\gamma_c v_c}
\left[
\frac{3}2\ln x
+\cst'
+3\sqrt{\frac{2\pi v_c}{\gamma_c^3\,v''(\gamma_c)}}x^{-1/2}+\cdots
\right].
\label{tx5}
\end{equation}
The main advantage of (\ref{tx1}-\ref{tx5}) over
(\ref{Xt1}-\ref{Xt4},\ref{EbSa}) is that they  still make sense when space
is discrete (with a real constant $\cst'$, not a periodic function of
time). We will see that they remain valid for our lattice
model~\eqref{evolution}.

\section{ The key formula for the position of the front}
%=======================================================
\label{S:key}
In this section we consider the front $h_n(t)$ defined by~\eqref{evolution}
and we establish relation~\eqref{key-formula} between the
initial condition $h_n(0)$ and the first times $t_n$ at which
$h_n(t)$ reaches the value $1$.
Here we limit our discussion to the case
$a >0 $
and to initial conditions of the form 
\begin{equation}
h_n(0)=
\begin{cases}
1 & \text{for $n \le 0$,} \\
 k_n & \text{for $n \ge 1$,}
\end{cases}
\label{initial-condition}
\end{equation}
where the $k_n$ are non-negative, smaller than 1 and non-increasing, i.e.
\begin{equation}
1> k_1\ge k_2\ge k_3\ge \cdots \ge 0.
\label{monotone}
\end{equation}
Clearly, as $a>0$, for a monotonic initial condition
\eqref{monotone}, the solution $h_n(t)$ of \eqref{evolution} remains
monotonic at any later time. One can define $t_n$ as the time when
$h_n(t)$ reaches 1 for the first time (i.e.\@ $h_n(t)=1$ for $t \ge t_n$
while $h_n(t) <1 $ for $t < t_n$). 
The monotonicity \eqref{monotone} of the initial condition implies the
monotonicity of the times $t_n$
\begin{equation}
0 < t_1 < t_2 < \cdots < t_n < \cdots
\end{equation}

Most of the properties of the solutions of \eqref{evolution} with the initial
conditions \eqref{initial-condition} discussed in this paper will be based
on the following exact formula
\begin{equation}
 \sum_{n=1}^\infty k_n \lambda^n = - \frac{a \lambda }{ 1 + a \lambda} 
+ \frac{a+1 }{ 1 + a \lambda} \sum_{n=1}^\infty e^{-(1+ a \lambda ) t_n } 
 \,\lambda^n ,
\label{key-formula}
\end{equation}
which relates the generating function of the initial condition $\{k_n\}$ to
the times $\{t_n \}$.
\bigbreak

Formula~\eqref{key-formula} can be derived as follows. 
If one defines the generating functions
\begin{equation}
H_m(t) = \sum_{n \ge m} h_n(t)  \lambda^{n-m+1},
\end{equation}
one can see that for $t_{m-1} \le t \le t_m$ (with the convention that $t_0=0$) the evolution of 
$H_m(t)$ is given by
\begin{equation}
\frac{\diffd H_m(t) }{ \diffd t} = (1+ a \lambda) H_m(t) +a \lambda  .
\end{equation}
This of course can be easily solved to give
\begin{equation}
H_m(t)= - \frac{a\lambda }{ 1 + a \lambda} + \Phi_m e^{ (1 + a \lambda)t},
\end{equation}
where the $\Phi_m$'s are constants of integration. These $\Phi_m$'s can be
determined by matching the solutions at times $0$, $t_1$, $t_2$, \ldots:
\begin{equation}
H_1(0)= \sum_{n=1}^\infty k_n  \lambda^n,\qquad
H_m(t_m)= \lambda \big( 1 + H_{m+1} (t_m)\big),
\end{equation}
and one gets that
for $t_{m-1} \le t \le t_m$

\begin{equation}
H_m(t)=- \frac{a \lambda}{ 1 + a \lambda} +\frac{a+1}{1 + a \lambda} 
\sum_{n = m}^\infty e^{(1+ a \lambda ) (t-t_n) } \, \lambda^{n+1-m} .
\label{all-time}
\end{equation}
Then \eqref{key-formula} follows
 by taking $m=1$ and $t=0$ in \eqref{all-time}.

\bigbreak

Remark that
formula~\eqref{key-formula} appears as the solution of a kind of inverse
problem: given the times $t_n$, one can compute the initial profile $k_n$
by expanding in powers of $\lambda$. This gives expressions
of $k_n$ in terms of the times $t_m$'s for $m \le n$. Alternatively one can
determine the times $t_n$ in terms of the initial profile $k_m$ for $m \le
n$: 
\begin{equation}
e^{-t_1}=\frac{a+k_1}{a+1},\quad
e^{-t_2}=\frac{a k_1+k_2}{a+1}+a t_1 e^{-t_1},\quad
e^{-t_3}=\frac{a k_2+k_3}{a+1}+a t_2 e^{-t_2}-\frac{(a t_1)^2}2e^{-t_1},
\end{equation}
etc. 
Unfortunately these expressions become quickly too complicated to allow to
determine how the times $t_n$ depend asymptotically on the initial profile
$\{k_n\}$ for large $n$. How these asymptotics can be understood from
\eqref{key-formula} will be discussed in Section~\ref{S:position}.

\section{Travelling wave solutions}
%=================================
\label{S:TW}
\subsection{The exact shape of the travelling waves}
%--------------------------------------------------

As usual, with travelling wave equations, the first solutions one can try
to determine are travelling wave solutions moving at a certain velocity
$v$. Because the $h_n (t)$ are defined on a lattice, a travelling wave
solution moving at velocity $v$ satisfies
\begin{equation}
h_n(t) 
= h_{n+1} \Big(t + \frac1v \Big).
\label{hW}
\end{equation}
Clearly this implies that
the times $t_n$ form an arithmetic progression, and by shifting the origin
of time, one can choose
\begin{equation}
t_n = \frac nv.
\end{equation}
This immediately gives, using \eqref{all-time}, the generating
function of the front shape at all times: for example for $0 \le t \le
t_1=1/v$,
one takes $m=1$ in \eqref{all-time} and gets
\begin{equation}
\sum_{n \ge 1} h_n(t) \lambda^n = 
- \frac{a \lambda }{1 + a \lambda}
+\frac{a+1}{1+a\lambda}\times
\frac{\lambda e^{(1+ a \lambda ) t } }{e^{ (1+ a \lambda )/v } - \lambda}.
\label{generat}
\end{equation}

Another way of determining the travelling wave solutions is to look
directly for solutions of~\eqref{evolution} of the form~\eqref{hW}. One
sees that $W_v$ must satisfy
\begin{equation}
W_v(x)=1\text{ for $x\le 0$},\qquad
W_v(x)+a W_v(x-1)+v W_v'(x)=0\text{ for $x>0$}.
\label{omegav}
\end{equation}
These equations can be solved iteratively: for $x\le 0$, one already knows
that $ W_v(x)=1$. For $x\in[0,1]$ one has therefore
$ W_v+v W_v'+a=0$, which implies for ($x\in[0,1]$) that 
$ W_v(x)=(a+1)e^{-x/v}-a$ (the integration
constant being fixed by continuity at $x=0$). Knowing $W_v(x) $ for
$x\in[0,1]$, one can solve \eqref{omegav} for $x\in[1,2]$ and so on.
\begin{equation}
 W_v(x)=\begin{cases}
1&\text{if $x\le 0$},\\
(a+1)e^{-x/v}-a&\text{if $x\in[0,1]$},\\
\frac{a(1+a)}v(1-v-x)e^{(1-x)/v}+(1+a)e^{-x/v}+a^2&\text{if $x\in[1,2]$},\\
\ldots
\end{cases}
\label{exactomega}
\end{equation}
In fact one can solve directly~\eqref{omegav} by considering for
$x\in[0,1]$ the generating
function $(\lambda,x)\mapsto\sum_n \lambda^n W_n(x+n)$. Then, as can be checked
directly from~\eqref{generat} and~\eqref{omegav}, $W_v(x)$ and $h_n(t)$ are
related for all $n\in\Z$ and $t\ge0$ by
\begin{equation}
h_n(t)=W_v(n-vt).
\end{equation}

\subsection{The decay of the travelling waves}
%---------------------------------------------

The large $n$ behavior of the travelling wave $h_n(t)$ (or equivalently the
behavior of $W_v(x)$ for large~$x$) can be understood by analyzing the
singularities in $\lambda$ of the right hand side of~\eqref{generat}.
These singularities
are poles located at all the real or complex zeros of 
\begin{equation}
 e^{ (1+ a \lambda )/ v } - \lambda=0.
\label{singularities}
\end{equation}
(one checks there is no pole at
$\lambda=-1/a$)
and each pole gives rise to an exponential decay in $h_n(t)$. Using
$\lambda=\exp(\gamma)$, \eqref{singularities} can be rewritten into
\begin{equation}
 v(\gamma)= \frac{1+ a e^\gamma}{\gamma} .
\label{singularities-bis}
\end{equation}
which is the dispersion relation for~\eqref{evolution},
similar to~\eqref{v-KPP} or to~\eqref{new-v}.
In fact, one can obtain \eqref{singularities-bis} as in
Section~\ref{S:recall} by looking for the velocity $v$ compatible in
\eqref{omegav} with an exponentially decaying travelling wave of the form
$W_v(x) \sim e^{-\gamma x}$.

One is then led to distinguish three cases depending on the number of real
solutions of \eqref{singularities-bis}. There is a critical value $v_c$ where
\eqref{singularities-bis} has a double zero on the real axis. This critical
velocity $v_c$ and the corresponding decay rate $\gamma_c$ are the solution
of \begin{equation}
a (\gamma_c -1) e^{\gamma_c} =1, \qquad
v_c = \frac{1}{\gamma_c-1}.
\label{omegac}
\end{equation}

\begin{enumerate}
\item For $v < v_c$, there is no real $\lambda$ solution of
\eqref{singularities},  but there are complex roots. The large $n$ behavior of $h_n(t)$ is then governed by the two roots
$\lambda_1$ and $\lambda_1^*$ of \eqref{singularities} closest to the origin
\begin{equation}
h_n(0) \simeq \frac{ (a+1) v }{ (1+ a \lambda_1) (v - a \lambda_1)} 
\lambda_1^{ - n}  +  \text{c.c.} 
\label{large-n}
\end{equation}
Because $\lambda_1$ and $\lambda_1^*$ are complex, $h_n(t)$ changes its
sign as $n$ varies. So for $v < v_c$, as in the Fisher-KPP case,
there are travelling wave solutions, but they
fail to give positive profiles. 
\item For $v=v_c$ given by \eqref{omegac}, there is a double real root
$\lambda_c = e^{\gamma_c}$ of \eqref{singularities}.
Then for large $n$ the profile is of the form
\begin{equation}
h_n(0) \simeq \frac{2 (1+a)}{1+ v_c} \left[n + \frac{1 + 4 v_c
}{3( 1 + v_c)} \right] e^{-\gamma_c n}.
\label{large-n2}
\end{equation}

\item For $v > v_c $ there are two real roots $1<\lambda_1< \lambda_2$
 of
\eqref{singularities}
and the large $n$ behavior is controlled by the smallest root:
\begin{equation}
h_n(0) \simeq \frac{ (a+1)v }{(1+ a \lambda_1) (v - a \lambda_1)}
\lambda_1^{- n} .
\label{large-n3}
\end{equation}
\end{enumerate}
We see that for all velocities, we get \emph{explicit} expressions of the
prefactors of the exponential decay of the travelling waves. These
prefactors are in general not known for more traditional travelling wave
equations, such as the Fisher-KPP equation \eqref{FKPP}.

\begin{figure}[ht]
\centering
\includegraphics[width=.9\textwidth]{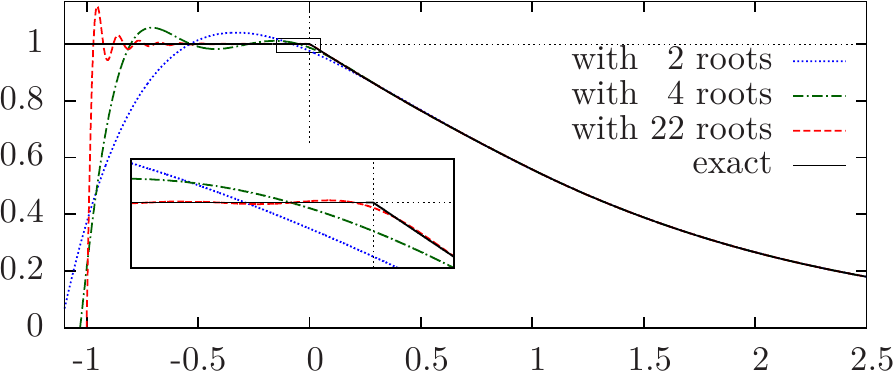}
\caption{The travelling wave $W_v(x)$ solution of \eqref{omegav} for $v=4$
and $a=1$ as a function of $x$. The plain line labeled ``exact'' is the
exact small-$x$ solution~\eqref{exactomega}. The dashed lines are the sums
\eqref{large-n1} truncated to a given number of first terms: ``with
2 roots'' means only the two real roots
$\lambda_1$ and $\lambda_2$, ``with 4 roots'' means the two real roots and
the first pair of complex conjugate roots and ``with 22 roots''
means the two real roots and the ten pairs of complex conjugates roots
closest to the origin. The inset is a zoom of the small rectangle around
$x=0$ and $W_v=1$.}
\label{fig1}
\end{figure}

In each case, corrections to~(\ref{large-n}-\ref{large-n3}) can be
obtained from the contributions of the other roots of
\eqref{singularities}. For instance, in the cases $v<v_c$ or $v>v_c$ one
could write
\begin{equation}
h_n(0) \simeq \sum_r \frac{ (a+1) v }{ (1+ a \lambda_r) (v - a \lambda_r)}
\lambda_r^{ - n} ,
\label{large-n1}
\end{equation}
where the sum is over all the complex roots $\lambda_r$ of
\eqref{singularities}.
In Figure~\ref{fig1}, we compare the exact solution \eqref{exactomega} of
\eqref{omegav} with the asymptotic expansion \eqref{large-n1} truncated to
a finite number of roots of \eqref{singularities} closest to the origin and
one can see that the truncation gives a very good fit of the actual
solution. 

We have seen that the travelling waves for $v<v_c$ were oscillatory. For
$v\ge v_c$, they decrease monotonically towards~0; this 
can be seen directly
from equation~\eqref{omegav} verified by $W_v(x)$:  write
$W_v(x)=R(x)e^{-\gamma x}$ with $\gamma$ a real positive number related to
$v$ through the dispersion relation \eqref{singularities-bis}.
(Notice that choosing such a $\gamma$ is impossible if $v<v_c$.) Then
\eqref{omegav} gives
\begin{equation}
R(x)=e^{\gamma x}\text{ for $x\le 0$},\qquad
a e^{\gamma} \big[R(x-1)- R(x)\big] +v R'(x)=0\text{ for $x>0$}.
\label{eqR}
\end{equation}
As $R(x)$ is strictly increasing for $x<0$ it must be strictly increasing
for all reals; otherwise, on the first local maximum $x_m$, one would have
$R'(x_m)=0$ and $R(x_m)>R(x_m-1)$  which is incompatible with~\eqref{eqR}.
Hence, $W_v(x)$ is positive and, from~\eqref{omegav}, strictly decreasing.

\section{How the initial condition determines the asymptotic regime}
%===================================================================
\label{S:position}

We now discuss how the position of the front at large times (or
equivalently the large~$n$ asymptotics of the times $t_n$) depends on
the initial condition.

First, by using mostly a comparison property, we will show that the final velocity
of the front is determined by the large~$n$ decay of the
initial condition $k_n$. Then, we will recover the logarithmic corrections
(\ref{Xt1}-\ref{Xt4}) and sub-leading terms as in \eqref{EbSa} by analyzing
the key relation \eqref{key-formula}
between the initial profile $k_n$ and the times $t_n$.

We write
\eqref{key-formula} as
\begin{equation}
(1+ a \lambda) K(\lambda)= -a \lambda + (a+1) T(\lambda),
\label{2-functions}
\end{equation}
where the two functions $K(\lambda) $ and $T(\lambda)$ are defined by
\begin{equation}
K(\lambda)= \sum_{n=1}^\infty k_n  \lambda^n,
\qquad
T(\lambda) = \sum_{n=1}^\infty e^{-(1+ a \lambda ) t_n }  \lambda^n  .
\label{2-functions-def}
\end{equation}
The large $n$ behavior of the $k_n$'s and of the $t_n$'s determines
the domain of convergence of these two sums and one
can try to use \eqref{2-functions} to relate their
singularities. 

When $\lambda=e^\gamma>1$, we will often use the following form of
$T(\lambda)$ written in terms of the dispersion relation $v(\gamma)$:
\begin{equation}
T(e^\gamma) = \sum_{n=1}^\infty e^{\gamma [n - v(\gamma)
{t_n}]}.
\label{T2}
\end{equation}

\subsection{Selection of the velocity}
%-------------------------------------

Let us first show that the
final velocity of the front is determined by the large~$n$ behavior of
the initial condition~$k_n$ in the same way as for other equations in the
Fisher-KPP class.
To do this, we use an obvious comparison
property; considering two initial conditions $\{k_n^{(1)}\}$ and
$\{k_n^{(2)}\}$ with the corresponding times $\{t_n^{(1)}\}$ and
$\{t_n^{(2)}\}$, one has
\begin{equation}
\text{if $0\le k_n^{(1)}\le k_n^{(2)}$ for all $n$, then $t_n^{(1)}\ge
t_n^{(2)}$ for all $n$.}
\end{equation}
To keep the discussion simple, we focus only on initial conditions $\{k_n\}$
with $k_n\ge0$ and the following simple asymptotics:
\begin{itemize}
\item If $k_n\sim n^\alpha e^{-\gamma n}$ with $0<\gamma<\gamma_c$.\\Pick an
$\epsilon>0$ small enough so that $0<\gamma-\epsilon$ and
$\gamma+\epsilon<\gamma_c$,
and consider the two travelling waves going at velocities
$v(\gamma-\epsilon)$  and $v(\gamma+\epsilon)$ (they decay respectively
like $e^{-(\gamma-\epsilon)n}$) and
 $e^{-(\gamma+\epsilon)n}$). It is
clear that the initial condition $\{k_n\}$ can be sandwiched between these
two travelling waves suitably shifted in space, so that, by using the
comparison property one gets
\begin{equation}
\frac1{v(\gamma-\epsilon)}\le \liminf_{n\to\infty} \frac
{t_n}n\le\limsup_{n\to\infty}\frac{t_n}n\le\frac1{v(\gamma+\epsilon)}.
\end{equation}
Now take $\epsilon\to0$ to get
\begin{equation}
\lim_{n\to\infty}\frac{t_n}n=\frac1{v(\gamma)}.
\end{equation}
\item If $k_n=0$.\\It takes a time~$t_n$ to have $h_n(t)=1$. But at
time~$t_n$, the $h_{n+m}(t)$ for $m>0$ are positive so that, from the
comparison property, one has
\begin{equation}
t_{n+m}\le t_n+t_m.
\end{equation}
The sequence $\{t_n\}$ is sub-additive and therefore $t_n/n$ has
a limit which we call $1/v$. By comparing the initial profile $k_n=0$ to
the travelling wave going at velocity $v_c$, one must have $1/v\ge1/v_c$.

We are now going to show that $1/v$ cannot be strictly
greater than $1/v_c$. Indeed, if we had $1/v>1/v_c$,
the series~\eqref{T2} defining $T(\lambda)$ would be uniformly convergent
on the whole positive real axis $\lambda$ because $v(\gamma)t_n/n$ would
eventually be larger than $1+\epsilon$ for some $\epsilon>0$.
One would then get
\begin{equation}
T'(\lambda)=\sum_{n\ge1} \lambda^n e^{-(1+a \lambda )t_n} \Big[
\frac{ n}{\lambda} - a t_n \Big]
\qquad\text{for all $\lambda\ge0$}
\label{derser}
\end{equation}
because the series~\eqref{derser} would also be uniformly convergent. 

However, for real and large enough $\lambda$ (at least for $\lambda
> \min_n n / (a t_n)$), one would obtain $T'(\lambda)<0$. But, with
$k_n=0$ one has $K(\lambda)=0$ and from \eqref{2-functions}
$T'(\lambda)=a/(a+1)$, in contradiction with  $T'(\lambda)<0$.

We conclude that one must have
\begin{equation}
\lim_{n\to\infty} \frac{t_n} n=\frac1{v_c}\qquad\text{if $k_n=0$ for
$n\ge1$.}
\end{equation}

\item If $k_n\sim n^\alpha e^{-\gamma_c n}$ or if
$k_n=o\big(e^{-\gamma_cn}\big)$.\\Again, by the comparison property, the
initial condition can be sandwiched between $k_n=0$ and, for any
$\epsilon>0$, the suitably shifted travelling wave going at velocity
$v(\gamma_c-\epsilon)$. This leads to
conclude that
\begin{equation}
\lim_{n\to\infty} \frac {t_n}n =\frac1{v_c}.
\end{equation}
\end{itemize}
The velocity selection thus works as for other equations of the Fisher-KPP
class.

\subsection{Sub-leading corrections}
%-----------------------------------
We limit the discussion to initial conditions similar to those discussed
in the previous section which lead to a front with
some asymptotic velocity~$V$:
\begin{equation}
\lim_{n\to\infty}
\frac{t_n}{ n} = \frac{1 }{ V}.
\label{hasalimit}
\end{equation} 
We also assume that $k_n\ge0$ which implies that $V\ge v_c$ as was shown in
the previous section.

If $V>v_c$, write $V=v(\gamma_1)=v(\gamma_2)$ with
$0<\gamma_1<\gamma_c<\gamma_2$. We have seen that this velocity is reached 
for initial conditions such as $k_n\sim n^\alpha  e^{-\gamma_1 n}$.
In \eqref{T2}, it is then clear that the series $T(\lambda)$ is
divergent for $\lambda\in(e^{\gamma_1},e^{\gamma_2})$ and convergent for $
\lambda<e^{\gamma_1}$ or $\lambda>e^{\gamma_2}$. Furthermore,
in~\eqref{2-functions-def}, the radius of convergence of
$K(\lambda)$ is $e^{\gamma_1}$ and, as $k_n>0$, the function $K(\lambda)$
must have a singularity at $\lambda=e^{\gamma_1}$. We thus see that both
$T(\lambda)$ and $K(\lambda)$ become singular as $\lambda$
approaches $e^{\gamma_1}$ from below on the real axis.
By matching the singularities of these two functions, we
will obtain the sub-leading corrections to
$t_n$ for large~$n$.

For $V=v_c$, if the initial condition is $k_n\sim
n^\alpha e^{-\gamma_c n}$, the same argument applies: both $T(\lambda)$ and
$K(\lambda)$ are singular when $\lambda$ reaches $e^{\gamma_c}$, and one
must match the singularities. But, with $V=v_c$, one could also have an
initial condition which decays faster than $e^{-\gamma_c n}$ and for
which the radius of convergence is larger than $e^{\gamma_c}$ (even,
possibly, infinite). Then, of course, $K(\lambda)$ would have no singularity at
$\lambda=e^{\gamma_c}$, even though the convergence of $T(\lambda)$ would remain
problematic when $\lambda$ approaches $e^{\gamma_c}$. We will see that the
large $n$ behavior of $t_n$ is tuned to ``erase'' the
singularities in $T(\lambda)$ at $e^{\gamma_c}$ to
satisfy~\eqref{2-functions}.

We  attack
the  problem by assuming that the $t_n$ are given and we try to obtain the
asymptotics of the $k_n$.
The starting point is thus to assume a velocity $V=v(\gamma_1)$ with
$\gamma_1\le\gamma_c$, and study $T(\lambda)$ when $\lambda$ gets close to
$e^{\gamma_1}$.
If one chooses,
in all generality,
\begin{equation}
t_n=\frac{n} V +\frac{\delta_n}{\gamma_1 V},
\label{defdelta}
\end{equation}
where $\delta_n/n\to0$, one gets from~\eqref{T2}
\begin{equation}
T(e^\gamma)=\sum_{n=1}^\infty
e^{\gamma\big[1-\frac{v(\gamma)}V\big]n-\frac{\gamma v(\gamma)}{\gamma_1
V}\delta_n}.
\end{equation}

Now we want to take $\gamma=\gamma_1-\epsilon$ and expand for small
$\epsilon$ in order to extract the nature of the singularity.
Two cases arise:
\begin{itemize}
\item If $V>v_c$ (which means $\gamma_1<\gamma_c$),
then $v'(\gamma_1)<0$ and to leading order
\begin{equation}
T(e^{\gamma_1-\epsilon})=\sum_{n=1}^\infty
\exp\Big[\Big(\frac{\gamma_1 v'(\gamma_1)}V\epsilon+\cdots\Big)n
-(1-\mu\epsilon+\cdots)
\delta_n\Big]
\quad\text{for $V>v_c$},
\label{g1<gc}
\end{equation}
with $\mu=1/\gamma_1+v'(\gamma_1)/V$.

\item If $V=v_c$ (which means $\gamma_1=\gamma_c$), then $v'(\gamma_c)=0$ and one must push the
expansion further:
\begin{equation}
T(e^{\gamma_c-\epsilon})=\sum_{n=1}^\infty
\exp\Big[\Big(-
\frac{\gamma_cv''(\gamma_c)}{2v_c}\epsilon^2+\cdots\Big)n
-\Big(1-\frac1{\gamma_c}\epsilon+\cdots\Big) \delta_n\Big]
\quad\text{for $V=v_c$}.
\label{g1=gc}
\end{equation}
\end{itemize}

It is already clear that cases $V>v_c$ and $V=v_c$ need to be discussed
separately.
Equations~\eqref{g1<gc} and~\eqref{g1=gc} are the starting point of our
analysis which is presented in detail in the following subsections.

We will
make heavy use of the following formulas: for $\epsilon>0$ small,
\begin{align}
\sum_{n\ge1} n^\alpha e^{-\epsilon n}\Bigg|_\text{singular}
&=
\begin{cases}
\displaystyle
\frac{\Gamma(1+\alpha)}{\epsilon^{1+\alpha}}&\text{if $\alpha$ is not
a negative integer},\\[2ex]
\displaystyle
\frac{(-1)^\alpha\epsilon^{-\alpha-1}\ln\epsilon}{(-\alpha-1)!}
&
\text{if $\alpha$ is a negative integer}.
\end{cases}
\label{divergence}
\\[1ex]
\sum_{n\ge1} (\ln n) n^\alpha e^{-\epsilon n}\Bigg|_\text{singular}
&=
\begin{cases}
\displaystyle
\frac{
-\Gamma(1+\alpha)\ln\epsilon+\mathcal O(1)}{\epsilon^{1+\alpha}}&\text{if $\alpha$ is not a negative integer},\\[2ex]
\displaystyle
\frac{(-\epsilon)^{-\alpha-1}}{(-\alpha-1)!}
\left[\frac{\ln^2\epsilon}2+
\mathcal O(\ln\epsilon)\right]
&
\text{if $\alpha$ is a negative integer},
\end{cases}
\label{divergence2}
\end{align}
where the meaning of ``singular'' for a function $F(\epsilon)$ with
a singularity at $0$ is that the difference between
$F(\epsilon)$ and $F(\epsilon)\big|_\text{singular}$ is a  regular
function of $\epsilon$ which can be expanded as a power series.

\subsubsection{For $V> v_c$}
%---------------------------
As explained above we write $V=v(\gamma_1)$ with $\gamma_1<\gamma_c$, and
we choose $t_n$ such that $t_n/n$ that converges to $1/V$. If one chooses
\begin{equation}
t_n = \frac nV + \frac{B \ln n + C }{\gamma_1 V},
\label{tn2}
\end{equation}
by keeping the leading order in \eqref{g1<gc} and using~\eqref{divergence}
one gets for $B$ not a positive integer
\begin{equation}
T(e^{\gamma_1-\epsilon})\Big|_\text{singular}
\simeq
\Gamma(1-B) e^{-C}\Big(\frac V {-v'(\gamma_1)\gamma_1
\epsilon}\Big)^{1-B}.
\label{Tn2}
\end{equation} 
It is then easy to check that matching the singularities leads to an
initial condition decaying as
\begin{equation}
k_n \simeq  \frac{(1+a)e^{-C} }{-v'(\gamma_1)
 \gamma_1^2}\left[\frac{Vn}{-\gamma_1 v'(\gamma_1)}\right]^{-B}
e^{-\gamma_1 n} . 
\label{kn2}
\end{equation}
\emph{Remarks:} As can be easily checked, even though~\eqref{Tn2} is not
valid if $B$ is a positive integer, \eqref{kn2} is. 
One can also check that for $B=C=0$ one recovers the asymptotics~\eqref{large-n3}
of the travelling wave.

\subsubsection{For $V= v_c$}
%------------------------

if $V= v_c$, the main difference with the previous case is that
$v(\gamma_c-\epsilon) - v(\gamma_c) \sim \epsilon^2$ as $\epsilon\to0$ and
one must use the expansion~\eqref{g1=gc} instead of~\eqref{g1<gc}.
As before, we choose a specific form for the times~$t_n$ which allow 
to easily make the comparison with the different cases
(\ref{Xt2}-\ref{Xt4}) of the Fisher-KPP equation:
\begin{equation}
 t_n = \frac{n }{ v_c} + \frac{B \ln n + C}{\gamma_c v_c} .
\label{case1}
\end{equation}
With $\delta_n=B\ln +C$ into \eqref{g1=gc}, one obtains generically
(when $B\not\in\{1,3/2,2,5/2,3,\ldots\}$,
see discussion below)
\begin{equation}
T(e^{\gamma_c-\epsilon}) \Big|_\text{singular} \simeq 
 e^{-C }\Gamma(1-B)\left(\frac{\gamma_c v''(\gamma_c)}{ 2  v_c}
\right)^{B -1}\epsilon^{2B-2}.
\label{sing1}
\end{equation}
Then, using \eqref{divergence} again and \eqref{2-functions}, one gets
\begin{equation}
 k_n \simeq \frac{a+1 }{\gamma_c  v_c} 
 e^{-C }\left(\frac{\gamma_c v''(\gamma_c)}{ 2  v_c}
\right)^{B -1} 
 \frac{ \Gamma\left(1-B \right) 
 }{ \Gamma\left(2-2 B \right)}
 n^{1-2 B }  e^{-\gamma_c n}  .
\label{case1-kn}
\end{equation}
We see that the asymptotics of the initial condition \eqref{case1-kn} and
of the times \eqref{case1} for large~$n$ are related as in the Fisher-KPP
case \eqref{tx3} and that the constant term in~\eqref{tx3} can be determined.
As in~\eqref{tn2}, one must be careful when $B$ is a positive integer:
\eqref{sing1} should be modified to include the logarithmic correction
of~\eqref{divergence}, but \eqref{case1-kn} is not modified as can easily
be checked (the ratio of the two Gamma functions has a limit). 

There is another difficulty when $B\in\{3/2, 5/2, 7/2, \ldots\}$:
for these values, the ratio of Gamma functions in~\eqref{case1-kn}
is zero. This means that an initial condition
$\{k_n\}$ leading to~\eqref{case1} with  $B=3/2$ (for instance) must
decrease faster than $n^{-2}e^{-\gamma_c n}$.
For these special values of $B$, the right hand side of \eqref{sing1}
is actually regular as $\epsilon^{2B-2}$ is a non-negative integer power of $\epsilon$;
any singular part of $T(e^{\gamma_c-\epsilon})$ must come from higher order
terms.

We are now going to show that no non-negative initial condition~$\{k_n\}$
can lead to a time sequence $\{t_n\}$ with an asymptotic expansion starting
as in \eqref{case1} with $B>3/2$.
To do so, we will show that the initial condition~$k_n=0$ leads to 
\eqref{case1} with $B=3/2$ (plus higher order corrections).
As any non-negative initial
condition must lead to times $\{t_n\}$ which are smaller than the times of
the $k_n=0$ initial condition, this will prove that $B$ cannot be larger than
$3/2$.

Consider therefore the case $k_n=0$; one has
$K(\lambda)=0$ and, from~\eqref{2-functions}, one gets
$T(\lambda)=a\lambda/(a+1)$. Obviously, $T(\lambda)$ has no singularity 
as $\lambda$ approaches $e^{\gamma_c}$, so the right hand
side of
\eqref{sing1} must be regular, which implies that
$B\in\{3/2,5/2,7/2,\ldots\}$.
We now rule out any value other than
$B=3/2$ by looking at the term of order $\epsilon$ in the expansion of
$T(e^{\gamma_c-\epsilon})$. One can check that the only terms of order
$\epsilon$ come from the $\epsilon\delta_n/\gamma_c$ in~\eqref{g1=gc} and
from~\eqref{sing1} if $B=3/2$, so one has
\begin{equation}
T(e^{\gamma_c-\epsilon})=T(e^{\gamma_c})+\left[\frac1{\gamma_c}\sum_{n=1}^\infty
\delta_ne^{-\delta_n}
+e^{-C}\Gamma(-1/2)\left(\frac{\gamma_c v''(\gamma_c)}{ 2  v_c}
\right)^{1/2}\one_{B=3/2}\right]\epsilon+o(\epsilon).
\label{orderepsilon}
\end{equation}
Notice \eqref{defdelta} that $\delta_n\ge0$ for the $k_n=0$ initial condition 
because it is below the travelling wave at velocity $v_c$ for which
$\delta_n=0$. The first term of order $\epsilon$ in \eqref{orderepsilon} is
therefore positive; on the other hand, the second term (only if $B=3/2$) is
negative. But, from $T(\lambda)=a\lambda/(a+1)$ the term of order
$\epsilon$ must be negative; therefore one must have $B=3/2$ for the zero
initial condition and, therefore, $B\le3/2$ for any non-negative initial
condition.

To summarize, the relationship between the times~\eqref{case1} and
the initial condition~\eqref{case1-kn} we have established in this section
is valid only for $B<3/2$ because we only consider non-negative initial
conditions. Furthermore, to have \eqref{case1} with $B=3/2$, one must
have an initial condition decreasing faster than
the $n^{-2}e^{-\gamma_c n}$ suggested by~\eqref{case1-kn}.
No non-negative initial condition can lead to~\eqref{case1} with $B>3/2$.

\subsubsection{For $V= v_c$ and $B=3/2$}
%------------------------

The case $B=3/2$ 
is of course the most delicate and it corresponds to
(\ref{Xt2},\ref{Xt4},\ref{EbSa}) in the Fisher-KPP case.
For the $t_n$ given by \eqref{case1} the leading singularity is 
not~\eqref{sing1} but rather
\begin{equation}
 T(e^{\gamma_c-\epsilon}) \Big|_\text{singular} \simeq 
3  e^{-C } 
 \sqrt{\frac{2 \pi v''(\gamma_c) }{ \gamma_c  v_c} }
\epsilon^2\ln\epsilon.
\label{sing2}
\end{equation}
(This term comes from the first order expansion of the
$\epsilon\delta_n/\gamma_c$ term in~\eqref{g1=gc}.)
Relating this to the $\{k_n\}$ through \eqref{2-functions}, it leads
through
\eqref{divergence} to $k_n\sim n^{-3}e^{-\gamma_c n}$ with a negative
prefactor. So there is no way for a non-negative
initial condition to be compatible with exactly~\eqref{case1}, without any
extra term.

Therefore, we need to add some corrections to~\eqref{case1} when
$B=3/2$.
Let us consider a correction of the form
\begin{equation}
 t_n = \frac n {v_c} + \frac{\frac32\ln n + C + D n^{-\xi}} 
{\gamma_c v_c}
\label{case2}
\end{equation}
for some $\xi>0$.
Plugging this correction into~\eqref{g1=gc} one gets
\begin{equation}
T(e^{\gamma_c-\epsilon})=\sum_{n=1}^\infty
e^{-C- \frac{\gamma_cv''(\gamma_c)}{2v_c}\epsilon^2n }\,
n^{-\frac32}\Big[1+\frac{3\epsilon}{2\gamma_c}\ln n-Dn^{-\xi}+\mathcal
\cdots\Big],
\end{equation}
where the ``$\cdots$'' contains smaller order terms of orders $n\epsilon^3$,
$n^{-2\xi}$, $\epsilon n^{-\xi}$, $\epsilon^2\ln^2n$, etc.
Consider in turns the terms in the square bracket. The ``$1$'' leads to
the right hand side of~\eqref{sing1} with $B=3/2$, which is simply a
regular term
linear in~$\epsilon$. The term in $\epsilon\ln n$ gives the right hand side
of~\eqref{sing2} and the $-Dn^{-\xi}$ contribution can be computed from
\begin{equation}
\sum_{n=1}^\infty
\left.e^{- \frac{\gamma_cv''(\gamma_c)}{2v_c}\epsilon^2n }\,
n^{-\frac32-\xi}\right|_\text{singular}
=
\begin{cases}
\Gamma(-\frac12-\xi)
\left(\frac{\gamma_cv''(\gamma_c)}{2v_c}\right)
	^{\frac12+\xi}\epsilon^{1+2\xi}
&\text{if $\xi\not\in\{\frac12,\frac32,\frac52,\ldots\}$},\\[2ex]
2\frac{\gamma_cv''(\gamma_c)}{2v_c}\epsilon^2\ln\epsilon
&\text{if $\xi=\frac12$}.
\end{cases}
\label{case2nxi}
\end{equation}
Several subcases must be considered
\begin{itemize}
\item
If $\xi>1/2$ this is smaller than $\epsilon^2\ln\epsilon$; therefore the
leading singularity is still given by~\eqref{sing2} which is incompatible
with a non-negative initial condition.
\item 
If $0<\xi<1/2$ the leading singularity for $T(e^{\gamma_c-\epsilon})$
is $\epsilon^{1+2\xi}$ as given by \eqref{case2nxi}. This leads to
\begin{equation}
k_n \simeq
-De^{-C}\ \frac{1+a}{\gamma_cv_c}\,
\frac{\Gamma(-\frac12-\xi)}
     {\Gamma(-1-2\xi)}
\left(\frac{\gamma_cv''(\gamma_c)}{2v_c}\right)
	^{\frac12+\xi}
n^{-2-2\xi}
e^{-n\gamma_c}.
\label{case2k}
\end{equation}
With $0<\xi<\frac12$, this is positive if $D>0$.

\item If $\xi=1/2$ the corrections from~\eqref{case2nxi} and
from~\eqref{sing2} are both of order~$\epsilon^2\ln\epsilon$. This leads to
\begin{equation}
k_n\simeq 2\frac{1+a}{\gamma_cv_c}  e^{-C } \left[
D \frac{\gamma_c v''(\gamma_c)}{v_c}
-
3\sqrt{\frac{2 \pi v''(\gamma_c) }{ \gamma_c  v_c} }
\,\right] n^{-3}e^{-\gamma_c n},
\label{case2x}
\end{equation}
which is positive if $D$ is large enough.
Notice also that the square bracket in~\eqref{case2x} vanishes for
\begin{equation}
D=3\sqrt{\frac{2\pi v_c}{\gamma_c^3v''(\gamma_c)}}.
\label{Dcrit}
\end{equation}
This means that initial conditions decaying faster than $n^{-3}e^{-\gamma_c
n}$ (including the zero initial condition) must lead to~\eqref{case2} with $\xi=1/2$ and $D$
given by~\eqref{Dcrit}. This is exactly the prediction~\eqref{tx5}.

\end{itemize}

To finish, notice that we found the first terms of the 
asymptotic expansion for the times $t_n$ when the initial condition
decays as $n^{\alpha}e^{-n\gamma_c}$ when $\alpha>-2$
(see~\eqref{case1} for $B<3/2$)
and when $\alpha<-2$ (it is of the form~\eqref{case2} with
$\xi=-1-\alpha/2$ for $-3<\alpha<-2$ and $\xi=1/2$ for
$\alpha\le-3$), but we did not yet considered the case where $k_n\simeq
n^{-2}e^{-n\gamma_c}$.
One can check that by taking, as in~\eqref{tx4},
\begin{equation}
 t_n = \frac n {v_c} + \frac{\frac32\ln n - \ln\ln n +C }  {\gamma_c
v_c} ,
\label{case4tn}
\end{equation}
one obtains
\begin{equation}
T(e^{\gamma_c-\epsilon}) \Big|_\text{singular} \simeq
e^{-C}\sqrt{\frac{8\pi\gamma_c v''(\gamma_c)}{v_c}}
\
\epsilon\ln\epsilon,
\label{sing4}
\end{equation}
which leads to
\begin{equation}
k_n\simeq\frac{1+a}{\gamma_c v_c}e^{-C}\sqrt{\frac{8\pi\gamma_c
v''(\gamma_c)}{v_c}}
\ n^{-2}e^{-\gamma_c n}.
\label{case4kn}
\end{equation}

\section{Summary}
\label{sec:summary}
%================
In the previous section, we have computed the initial conditions $k_n$
as a function of the times~$t_n$. Table~\ref{summary} summarizes
our results.

\begin{table}[ht]
\centering
\let\nl\\
\def\tabcolsep{4pt}
\def\arraystretch{2}
\begin{tabular}{|l|m{3.1cm}|m{7.8cm}|m{2.25cm}|}
\hline
1&
$\displaystyle k_n\sim n^\alpha e^{-\gamma n}$ \nl
with $\gamma<\gamma_c$ 
	&
$\displaystyle 
t_n\simeq \frac n {v(\gamma)} +\frac1{\gamma v(\gamma)}\Big[-\alpha\ln
n +C\Big]$
& see~(\ref{tn2},\ref{kn2})
\\[1ex] \hline
2&
$\displaystyle k_n\sim n^\alpha e^{-\gamma_c n}$  \nl
with $\alpha>-2$ 
&
$\displaystyle 
t_n\simeq \frac n {v_c} +\frac1{\gamma_c v_c}\left[\frac{1-\alpha}2\ln
n +C\right]$
& see~(\ref{case1},\ref{case1-kn})
\\[1ex] \hline
3&
$\displaystyle k_n\sim n^{-2} e^{-\gamma_c n}$
&
$\displaystyle 
t_n\simeq \frac n {v_c} +\frac1{\gamma_c v_c}\left[\frac{3}2\ln n -\ln\ln
n+C\right]$
& see~(\ref{case4tn},\ref{case4kn})
\\[1ex]
\hline
4&
$\displaystyle k_n\sim n^{\alpha} e^{-\gamma_c n}$\nl
with $-3\le\alpha<-2$
&
$\displaystyle 
t_n\simeq \frac n {v_c} +\frac1{\gamma_c v_c}\left[{\frac{3}2\ln n +C +D
n^{1+\frac\alpha2}}\right]$
& see~(\ref{case2},\ref{case2k},\ref{case2x})
\\[1ex]
\hline
5&
$\displaystyle k_n\ll n^{\alpha} e^{-\gamma_c n}$\nl
for some $\alpha<-3$
&
$\displaystyle
t_n\simeq \frac n {v_c} +\frac1{\gamma_c v_c}\left[{\frac{3}2\ln n +C +
3\sqrt{\frac{2\pi v_c}{\gamma_c^3v''(\gamma_c)}}
\,n^{-\frac12}}\right] $
&see~(\ref{case2},\ref{case2x},\ref{Dcrit})
\\[1ex] \hline
\end{tabular}
\caption{Asymptotics of $t_n$ as a function of the leading
behavior of the initial condition~$k_n$.}
\label{summary}
\end{table}

These asymptotics agree with all previously known results discussed in
Section~\ref{S:recall}. Case~4 is a new prediction, and the domain of
validity of Ebert-van Saarloos correction~\eqref{EbSa} from
\cite{EbertvanSaarloos.00} is made precise (case~5).

The constant $C$  can easily be computed in cases~1 to~3, but we
did not manage to get a closed expression in
cases~4 and~5. Similarly, we
have no expression for $D$  in case~4;
in particular, for $\alpha=-3$, the $D$ of case~4 is not
given by the prefactor of $n^{-1/2}$ in case~5 because for $\alpha=-3$ the
right hand side of~\eqref{case2x} must not vanish.

The vanishing terms~$n^{1+\alpha/2}$ and $n^{-1/2}$  in cases~4 and~5 depend only on
the leading behavior of $k_n$ for large~$n$. One could compute higher order
corrections in cases~1 to~3 using the same technique by
looking at the next singularities in $T(e^{\gamma-\epsilon})$, but one would need to know a bit more
about the asymptotic behavior of $k_n$: one would find that
\begin{multline}
\text{If }k_n= A n^\alpha e^{-\gamma
n}\left(1+o\Big(\frac{\ln n}n\Big)\right),\\
\text{then }t_n=\big[\text{as above}\big]+\begin{cases}
\displaystyle D\frac{\ln n}n&\text{in case~1 for $\alpha\ne0$},\\[2ex]
\displaystyle D\frac{\ln n}{\sqrt n}&\text{in case~2 for
$\alpha\not\in\{-1,0,1\}$},\\[2ex]
\displaystyle D\frac{1}{\sqrt n}&\text{in case~3},
\end{cases}
\label{sub}
\end{multline}
where the prefactor~$D$ could be computed in each case. These vanishing
corrections in cases~1 to~3 are less universal than in case~4 to~5 as they
do not depend \emph{only} on the leading  behavior of $k_n$ for large~$n$,
but also on the fact that the sub-leading behavior of $k_n$ decays fast
enough compared to the leading behavior.
For case~1 with $\alpha=0$ and case~2 with $\alpha=1$, the
initial condition behaves asymptotically as the travelling
wave eventually reached by the front, and vanishing corrections might depend
on the initial condition in a more complicated way.
Case 2 with $\alpha=-1$ or $\alpha=0$ are border
cases with slightly different corrections.

If we conjecture that the new results (cases 4 and 5) of Table~\ref{summary}
hold for the whole Fisher-KPP class one can obtain, by
inverting the relations between $t_n$ and $n$ of Table~\ref{summary},
the asymptotics of the position $X_t$ for initial conditions of the form
$u(x,0)\sim x^\alpha e^{-\gamma x}$. This is done in Table~\ref{summary2}.
\begin{table}[ht]
\centering
\let\nl\\
\def\arraystretch{2}
\begin{tabular}{|l|m{3.2cm}|m{7.8cm}|}
\hline
1&
$\displaystyle u(x,0)\sim x^\alpha e^{-\gamma x}$ \nl
with $\gamma<\gamma_c$ 
	&
$\displaystyle
X_t\simeq v(\gamma) t+\frac{\alpha}{\gamma}\ln t +C'$
\\[1ex] \hline
2&
$\displaystyle u(x,0)\sim x^\alpha e^{-\gamma_c x}$  \nl
with $\alpha>-2$ 
&
$\displaystyle X_t\simeq v_c t+\frac{\alpha-1}{2\gamma_c}\ln t +C'$
\\[1ex] \hline
3&
$\displaystyle u(x,0)\sim x^{-2} e^{-\gamma_c x}$
&
$\displaystyle X_t\simeq v_c t-\frac{3}{2\gamma_c}\ln t +
\frac1{\gamma_c}\ln\ln t+ C'$ 
\\[1ex]
\hline
4&
$\displaystyle u(x,0)\sim x^{\alpha} e^{-\gamma_c x}$\nl
with $-3\le\alpha<-2$
&
$\displaystyle X_t\simeq v_c t-\frac{3}{2\gamma_c}\ln t +
C'-D' t^{1+\frac\alpha2}$
\tabularnewline[1ex]
\hline
5&
$\displaystyle u(x,0)\ll x^{\alpha} e^{-\gamma_c x}$\nl
for some $\alpha<-3$
&
$\displaystyle X_t\simeq v_c t-\frac{3}{2\gamma_c}\ln t +
C'-3\sqrt{\frac{2\pi}{\gamma_c^5v''(\gamma_c)}}
\, t^{-\frac12}$
\\[1ex] \hline
\end{tabular}
\caption{Asymptotic expansion of $X_t$ as a function of the leading
behavior of the initial condition~$u(x,0)$.}
\label{summary2}
\end{table}

\section{Conclusion}
%===================

The main result of the present work is the exact relation
\eqref{key-formula} between the initial condition and the positions of the
front at time $t$ for the model \eqref{evolution}. Relating the asymptotics
of the $t_n$'s to those of the $k_n$'s, using the exact relation
\eqref{key-formula} is an interesting but not easy problem of complex
analysis. It allows to obtain
precise expressions of the shape of the travelling waves, including
prefactors which are usually not known in the usual equations of the
Fisher-KPP type. It also allows one to recover the known long time
asymptotics of the front position, and to get previously unknown
results; in particular, we have shown  how fast an initial condition should
decay to exhibit the Ebert-van Saarloos correction, and that there
is a range of initial conditions which exhibit
the $-3/2\ln t $ Bramson logarithmic term but for which the Ebert-van Saarloos
correction is modified.
(See cases~4 and~5 of
Tables~\ref{summary} and~\ref{summary2}.)

As shown here the analysis of the asymptotics \eqref{evolution}, using
complex analysis, is tedious but rather straightforward. Higher corrections
to the asymptotics of the
position could be determined. One could also try to study how, depending on the initial
condition, the asymptotic shape is reached. Furthermore, it would be interesting to
generalize \eqref{evolution} to evolutions involving more
than two neighboring sites, or to a non-lattice version of the model. 
More challenging would be to
 attack the noisy version of the
problem \cite{BDMM.06,MuellerMytnikQuastel.2010}.

\appendix

\section{An heuristic derivation of the positions of the front}
%========================================================

In this appendix we show that several expressions of the position of the
front for
a Fisher-KPP front can be recovered by considering a simplified version of the
Fisher-KPP equation~\eqref{FKPP} where the non-linear term is replaced by an
absorbing boundary.
Consider the following linearized Fisher-KPP equation with
a given time-dependent boundary~$X_t$ with $X_0=0$:
\begin{equation}
\begin{cases}
\displaystyle\frac{\partial u}{\partial t} =\frac{\partial^2 u}{\partial x^2}
+f'(0)u&\text{if $x>X_t$},\\[1ex]
\displaystyle u(X_t,t)=0.
\end{cases}
\label{bnd}
\end{equation}
For a given $y>0$, we look at the value $u(X_t+y,t)$ of the solution at
a distance $y$ from the boundary. Intuitively, if $X_t$ increases too
quickly with $t$, this quantity is pushed to zero. On the other hand, if
$X_t$ increases too slowly, it diverges with~$t$. It is only for
finely tuned choices of $X_t$ that $u(X_t+y,t)$ remains of order~1.

Now we suppose that $X_t$ is no longer given \textit{a priori} but is instead
determined by
\begin{equation}
u(X_t+1,t)=1.
\label{bnd2}
\end{equation}
It has been shown \cite{Henderson.2014} that the solution of
(\ref{bnd},\ref{bnd2}) for compactly supported initial conditions leads to
the same long time asymptotics for $X_t$ as for the Fisher-KPP equation
(see Section~\ref{S:recall}): one recovers the Bramson term~\eqref{Xt2} and
the Ebert-Van Saarloos correction~\eqref{EbSa}.

For initial conditions decaying fast enough, one expects $X_t$ to be
asymptotically linear. If $X_t$ were really linear (not only
asymptotically but at all times), \eqref{bnd} would be very easy to
solve. In this Appendix, we solve a simplified version
of~\eqref{bnd} where the boundary is replaced by a straight line. This
allows us to recover
the velocity and the logarithmic corrections (\ref{Xt1}-\ref{Xt4}) of the
Fisher-KPP equation.

The version we actually consider is therefore the following: For
each given time~$t$, we replace the boundary by a linear boundary of slope
$X_t/t$ and solve
\begin{equation}
\begin{cases}
\displaystyle\frac{\partial u}{\partial s} =\frac{\partial^2 u}{\partial x^2}
+f'(0)u&\text{if $x>\frac{X_t}t s$},\\[2ex]
\displaystyle u\Big(\frac{X_t}ts,s\Big)=0.
\end{cases}
\label{bnd3}
\end{equation}
We then tune the value of $X_t$ to satisfy \eqref{bnd2} at time~$t$.

For an initial condition $\delta(x-x_0)$ the solution to~\eqref{bnd3} is
\begin{equation}
g(x,s|x_0) = 
\frac{e^{f'(0)s} }{ \sqrt{4 \pi s}} \left[ 
\exp\left(-\frac{(x-x_0)^2 }{ 4 s }\right) 
-\exp\left( \frac{X_t}t x_0 -\frac {(x+x_0)^2 }{ 4 s }\right) \right].
\end{equation}
Taking $s=t$ and writing $x=X_t+y$, one obtains
\begin{equation}
g(X_t+y,t|x_0) =
\frac{1}{ \sqrt{4 \pi t}}
\exp\left[f'(0)t-\frac{(X_t+y)^2+x_0^2-2X_tx_0}{4t}\right]
2 \sinh\left(\frac{yx_0}{ 2 t }\right).
\end{equation}
Given a general initial condition $u(x_0,0)$ for $x_0>0$
one has
\begin{equation}
u(X_t+y,t) = \int_0^\infty \diffd x_0\,g(X_t+y,t|x_0) u(x_0,0) ,
\end{equation}
which, after writing $X_t = c t -\delta_t$ with $\delta_t\ll t$, leads to
\begin{equation}
\begin{aligned}
&u(X_t+y,t)=
\frac{1}{ \sqrt{\pi t}}
\exp\left[t\Big(f'(0)-\frac{c^2}4\Big)-\frac
c 2(y-\delta_t)-\frac{(y-\delta_t)^2}{4t}\right]\times I_t(y),
\\&\text{with}\quad
I_t(y)=\int_0^\infty\diffd x_0\,u(x_0,0)
\exp\left[\frac{c x_0}2-\frac{\delta_t x_0}{2t}-\frac{x_0^2}{4t}\right]
\sinh\Big(\frac{yx_0}{ 2 t }\Big).
\end{aligned}
\label{uI2}
\end{equation}

Depending on the initial condition $u(x_0,0)$, we can now determine for
which values of $c$ and $\delta_t$ the front $u(X_t+y,t)$ remains of
order~1 for $y$ of order~1 as $t$ increases.

\begin{itemize}
\item For $u(x_0,0) \simeq A e^{- \gamma x_0}$ with $\gamma<c/2$,\\
one finds that the integral $I_t(y)$ is dominated by $x_0 \simeq (c - 2 \gamma) t$.
One obtains
\begin{equation}
\begin{aligned}
&
I_t(y)\simeq A \sqrt{4\pi
t}\,\sinh\Big(\frac{c-2\gamma}2y\Big)\exp\left[\Big(\frac {c^2}4-\gamma
c +\gamma^2\Big)t
-\frac{c-2\gamma}2\delta_t
\right],
\\&\text{and}\
u(X_t+y,t)\simeq 2 A\sinh\Big[\frac{c-2\gamma}2y\Big]
\exp\left[\big(f'(0)-\gamma c +\gamma^2\big)t+\gamma\delta_t-\frac c2y\right].
\end{aligned}
\label{Iu1}
\end{equation}
Writing $u(X_t+y,t)\sim 1$ leads to $c=\gamma+f'(0)/\gamma=v(\gamma)$ and
$\delta_t\simeq\cst$. The starting hypothesis $\gamma<c/2$ then translates into
$\gamma<\gamma_c=\sqrt{f'(0)}$. We conclude that
\begin{equation}
\text{For }u(x_0,0)\sim e^{-\gamma x_0}\text{ with }\gamma<\gamma_c,\qquad
X_t\simeq v(\gamma)t +C,
\end{equation}
as in \eqref{Xt1}.

\item For $u(x_0,0) \simeq A x_0^\alpha e^{- \gamma x_0}$ with $\gamma<c/2$,\\
the integral $I_t(y)$  is again dominated by $x_0\simeq (c-2\gamma) t$. The large~$t$
expression of $u(X_t+y,t)$ has an extra term $[(c-2\gamma)t]^\alpha$ which
is canceled by taking now $\delta_t\simeq -\frac\alpha\gamma\ln t +\cst$. (The
value of $c$ remains the same.) We conclude that
\begin{equation}
\text{For }u(x_0,0)\sim x_0^\alpha e^{-\gamma x_0}\text{ with
}\gamma<\gamma_c,\qquad
X_t\simeq v(\gamma)t +\frac\alpha\gamma\ln t +C.
\end{equation}

\item For $u(x_0,0) \ll e^{- \gamma x_0}$ for some $\gamma>c/2$ (steep
initial condition),\\
the integral $I_t(y)$ is dominated by $x_0$ of order~1. This leads to
\begin{equation}
\begin{aligned}
&I_t(y)\simeq\frac y{2t} \int_0^\infty \diffd x_0\, u(x_0,0) x_0
\exp\big[\frac {c x_0}2\big],
\\&\text{and}\quad
u(X_t+y,t)
\sim
\frac{y}{ t^{3/2}}
\exp\left[t\Big(f'(0)-\frac{c^2}4\Big)-\frac
c 2(y-\delta_t)\right].
\end{aligned}
\label{Iu2}
\end{equation}
One needs to take $c=2\sqrt{f'(0)}=v_c=2\gamma_c$ and
$\delta_t=\frac{3}{2\gamma_c}\ln t + \cst$. The starting hypothesis
$\gamma>c/2$ translates into $\gamma>\gamma_c$ and we conclude that
\begin{equation}
\text{For }u(x_0,0)\ll  e^{-\gamma x_0}\text{ for some $\gamma>\gamma_c$},\qquad
X_t\simeq v_c t -\frac3{2\gamma_c}\ln t +C,
\label{Iu2x}
\end{equation}
as in \eqref{Xt2}.

\item For $u(x_0,0) \simeq A x_0^\alpha e^{- \frac c 2 x_0}$,\\
depending on the value of $\alpha$, the integral $I_t(y)$ is dominated by values of
$x_0$ of order~1 or of order $\sqrt t$. In any case, $x_0\ll t$ and one can
simplify $I_t(y)$ into
\begin{equation}
I_t(y)\simeq \frac y {2t}\int_0^\infty\diffd x_0\, u(x_0,0) x_0 
\exp\left[{\frac c2
x_0}-\frac{x_0^2}{4t}\right].
\end{equation}
When $\alpha < -2$, this integral is dominated by $x_0$ of order~1, the
Gaussian term can be dropped and one recovers \eqref{Iu2} and \eqref{Iu2x}.

When $\alpha\ge -2$, the integral is dominated by $x_0$ of order $\sqrt t$.
One gets
\begin{equation}
I_t(y)\simeq A\frac{y}{2t}\int_1^\infty \diffd
x_0\,x_0^{\alpha+1}\exp\left[-\frac{x_0^2}{4t}\right]
\simeq
\begin{cases}
\displaystyle A y 
2^{\alpha}\Gamma\Big(1+\frac\alpha2\Big)t^{\frac\alpha 2}
&\text{if $\alpha>-2$,}\\
\displaystyle A y \frac{\ln t}{4t}&\text{if $\alpha=-2$.}
\end{cases}
\end{equation}
Into \eqref{uI2} one must therefore take $c=2\sqrt{f'(0)}=v_c=2\gamma_c$ and
$\delta_t=\frac{1-\alpha}{2\gamma_c}\ln t+\cst$ if $\alpha>-2$ or
$\delta_t=\frac3         {2\gamma_c}\ln t-\frac1{\gamma_c}\ln\ln t$ if
$\alpha=-2$. We conclude that
\begin{equation}
\text{For }u(x_0,0)\sim x_0^\alpha e^{-\gamma_c x_0},\quad
X_t\simeq\begin{cases}
\displaystyle v_c t-\frac{3}{2\gamma_c}\ln t + C &\text{if
$\alpha<-2$},\\[2ex]
\displaystyle v_c t-\frac{3}{2\gamma_c}\ln t + \frac{\ln\ln t}{\gamma_c}+C
&\text{if $\alpha=-2$},\\[2ex]
\displaystyle v_c t-\frac{1-\alpha}{2\gamma_c}\ln t + C &\text{if
$\alpha>-2$},
\end{cases}
\end{equation}
as in (\ref{Xt2}-\ref{Xt4}).
\end{itemize}

\bibliography{journ,front,kpz}
\end{document}